\documentclass[journal]{IEEEtran}
\usepackage{graphicx}
\usepackage{cite}
\usepackage{picinpar}
\usepackage{amsmath}
\usepackage{url}
\usepackage{flushend}
\usepackage[latin1]{inputenc}
\usepackage{colortbl}
\usepackage{soul}
\usepackage{multirow}
\usepackage{pifont}
\usepackage{eqnarray}

\usepackage{todonotes}
\usepackage{color}
\usepackage{alltt}
\usepackage{blindtext}
\usepackage[hidelinks]{hyperref}
\usepackage{enumerate}
\usepackage{siunitx}
\usepackage{epstopdf}
\usepackage{pbox}
\usepackage{threeparttable}
\usepackage{algorithm2e}
\usepackage{tabularx}
\newcolumntype{Y}{>{\centering\arraybackslash}X}
\usepackage{xcolor}

\begin{document}
\title{Integrating Expert Knowledge with Domain Adaptation for Unsupervised Fault Diagnosis}
\author{
	\vskip 1em
	{
	Qin Wang,
	Cees Taal,
	Olga Fink
	}

	\thanks{
		
		{
		Qin Wang and Olga Fink are with Intelligent Maintenance Systems, ETH Zurich, Switzerland (e-mail: \{qwang, ofink\}@ethz.ch). Cees Taal is with SKF Group, Netherlands (email: cees.taal@skf.com).  This work was supported by the Swiss National Science Foundation (SNSF) Grant no. PP00P2\_176878. Our code and the synthetic dataset is available at \url{https://github.com/qinenergy/syn2real}.
		
		}
	}
}

\maketitle

\begin{abstract}
 Data-driven fault diagnosis methods often require abundant labeled examples for each fault type. On the contrary, real-world data is often unlabeled and consists of mostly healthy observations and only few samples of faulty conditions. The lack of labels and  fault samples imposes a significant challenge for existing data-driven fault diagnosis methods. In this paper, we aim to overcome this limitation by integrating expert knowledge with domain adaptation in a synthetic-to-real framework for unsupervised fault diagnosis. Motivated by the fact that domain experts often have a relatively good understanding on how different fault types affect healthy signals, in the first step of the proposed framework, a synthetic fault dataset is generated by augmenting real vibration samples of healthy bearings. This synthetic dataset  integrates expert knowledge and encodes class information about the faults types. However, models trained solely based on the synthetic data  often do not perform well because of the distinct distribution difference between the synthetically generated and  real faults.  To overcome this domain gap between the synthetic and real data, in the second step of the proposed framework, an imbalance-robust domain adaptation~(DA) approach is proposed to adapt the model from synthetic faults~(source) to the unlabeled real faults~(target) which suffer from severe class imbalance. The framework is evaluated on two unsupervised fault diagnosis cases for bearings, the CWRU laboratory dataset and a real-world wind-turbine dataset. Experimental results demonstrate that the generated faults are effective for encoding fault type information and the domain adaptation is robust against the different levels of class imbalance between faults.
\end{abstract}


{}

\definecolor{limegreen}{rgb}{0.2, 0.8, 0.2}
\definecolor{forestgreen}{rgb}{0.13, 0.55, 0.13}
\definecolor{greenhtml}{rgb}{0.0, 0.5, 0.0}

\section{Introduction}
Data-driven fault diagnosis methods often require a large number of labeled data to generalize well. Faults are, however, rare in real-world complex and safety critical systems. Therefore, a sufficient number of representative samples of faulty conditions is often impossible to be collected in real world applications. Recordings from industry assets often consist of a majority of healthy states and only few faults. In addition, not all fault types may have been captured by the different assets. Moreover, these recordings are often unlabeled because precisely identifying when and which fault is emerging can be difficult even for experienced domain experts. These real-world restrictions make learning fault patterns from unlabeled real-world data a challenging task.

One potential solution to overcome these challenges in this unsupervised fault diagnosis setup is to use synthetic faults as the supervision for the data-driven diagnosis models. For example, for bearing fault diagnosis, given operating conditions and bearing characteristics as input, synthetic vibration signals can be generated by 
highly accurate physical models, e.g, ~\cite{singh2015extensive}. By generating a large number of synthetic faults, a data-driven model can be trained solely based on the synthetic faults, and then evaluated on the real target data~\cite{zhang2001wavelet, eklund2006using, gryllias2012support, sobie2018simulation}. However, this way of using synthetic faults has several drawbacks. Firstly, detailed operating conditions can be unknown in order to achieve a realistic physical model. For example, in bearing vibration modeling we typically see that only the bearing is modeled ignoring surrounding rotating equipment, which can have a strong impact on the diagnosis performance. Secondly, even advanced simulations are not perfect, there will always be a distribution gap between the synthetic data and experimental measurements of mechanical systems~\cite{gao2020fem}. This domain gap between synthetic source and real target often leads to a significant performance degradation~\cite{ganin2014unsupervised} if the model is solely trained on the synthetic data. Thirdly, this pure synthetic data approach fails to make use of the available unlabeled real data, which can be potentially useful for providing additional information on the real faults. {Finally, in reality, developing accurate physical models for assets can be expensive and time consuming. For complex systems and complex physics of failure dynamics, detailed physics-based models may not even be available.}

To tackle the challenges of synthetic fault generation outlined above, we propose a novel framework for unsupervised fault diagnosis which relaxes the need of highly accurate physical models, while performing well on the real target data. {Motivated by the fact that domain experts often have a good understanding on how different fault types affect healthy signals, we propose to integrate} expert knowledge in synthetic data with imbalance-robust domain adaptation for unsupervised fault diagnosis. Unlike previous works~\cite{ gao2020fem} which use faults generated by highly accurate physical simulation models as the supervision, the proposed framework uses a relatively simple process {decided by the expert} for fault generation by augmenting healthy samples. 
To compensate on the potentially large domain gap between synthetic and real faults, the idea of domain adaptation~(DA) is adopted. We propose to align the conditional distributions between synthetic and real features. The proposed DA approach relies less on the quality of the synthetic fault simulator and is robust to the class imbalance in the target domain. Specifically, the proposed framework consists of two complementary parts. In the first stage, expert knowledge is used to generate synthetic data. In the second stage, DA is applied on the synthetic source and real target data to alleviate the domain gap under severe class imbalance.

In this work, we apply our framework to bearing fault diagnosis. More specifically, we rely on expert knowledge and use a relative simple approach where we synthesize vibration signals with fault-initiated pulse-trains corresponding to certain surface defects~\cite{singh2015extensive, randall2011}. We include more realistic vibration disturbances from surrounding rotating equipment, by mixing synthetic signals with healthy conditions. As a consequence, a reasonable number of samples can be generated for each fault, forming a balanced synthetic dataset. This synthetic dataset, thus makes use of expert knowledge, and can then facilitate the training of a data-driven fault diagnosis model. This proposed generation process relaxes the need of highly accurate physical models.

Although mixing healthy with synthetic defect signals will result in more realistic samples, a domain gap is still inevitable. To overcome this, the idea of unsupervised domain adaptation~\cite{fernando2013unsupervised,long2015learning, sun2015subspace } offers a potential solution. DA aims at learning a representative  model from the labeled synthetic source domain, at the same time leveraging unlabeled  data from the target domain to improve the model's generalization ability on the target data. Popular adversarial DA approaches~\cite{ganin2014unsupervised, tsai2018learning} use a domain discriminator to directly align the unconditional source and target distributions and minimize the discrepancy between them. They were originally designed and validated for image classification which often consists of balanced classes. 

\begin{figure}
    	\centering
	{\includegraphics[width=\columnwidth]{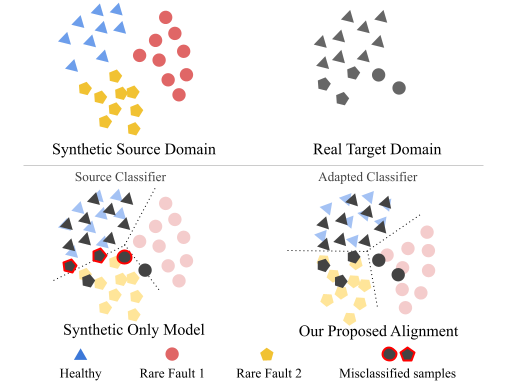}}
	\caption{The synthetic faults often suffer a distribution shift from the real faults, which leads to misclassified examples when models are trained only on the synthetic data. We propose to alleviate this problem by using domain adaptation to align the labeled synthetic domain with unlabeled target domain. The rare faults in the unlabeled target domain have very few samples, which makes the alignment harder. The proposed alignment method is robust to this class imbalance by aligning the conditional distributions. The model performance on the target domain can largely be improved by our approach.}
	\label{fig0}
	\vspace{-1mm}
\end{figure}

However, these standard adversarial DA methods make the implicit assumption that source and target domain share similar class distributions, and directly align the source and target unconditional data without using class information. While this assumption is often realistic for the original image datasets, this can be largely violated in the unsupervised fault diagnosis setups because of the nature of rare faults. Given an unlabeled dataset from real-world~(target) domain, its class distribution is most likely largely skewed towards the healthy class. Moreover, no more information can be inferred about the target class distribution or its level of class imbalance because the dataset is unlabeled. On the synthetic~(source) side, samples are usually generated for each fault class based on the healthy samples, leading to a class-balanced synthetic dataset. If we directly adopt the standard adversarial DA methods, the alignment is then performed between a class-balanced synthetic source domain and a highly imbalanced real-world target domain consisting of a majority of healthy data. This mismatch can further lead to a performance drop on the target data. On the one hand, the different faults in the source domain would mostly be aligned to the healthy state which dominates the target data. On the other hand, rare faults in the target data would be poorly adapted.

To address the challenge of imbalanced datasets in unsupervised fault diagnostics setup, we propose a novel imbalance-robust DA approach which overcomes the imbalance problem by making better use of available class information for the alignment. Specifically, we design the discriminator to align the distributions conditioned on classes instead of the unconditional ones. This is achieved by feeding the class information encoded by pseudo-labels to the discriminator. In addition, to make the training of the discriminator more stable and provide more training examples from rare faults, we propose to use a mixup~\cite{zhang2017mixup}-inspired augmentation to provide more support for the conditioned distributions on these rare classes. We illustrate the effect of the proposed DA approach in Figure~\ref{fig0}.

We demonstrate the effectiveness of our proposed framework on two different case studies. The first case study is based on the publicly available CWRU laboratory dataset. The second case study is based on the data collected from bearings of real wind turbine generators in the field. The proposed framework uses a relatively simple method for synthesizing the faults. However, it is still able to achieve a good performance with the help of the proposed DA approach. 

While the evaluation of the proposed framework was performed on two bearing datasets, the proposed synthetic-to-real framework can be easily implemented  for most industry assets. Moreover, the proposed imbalance-robust domain adaptation method can be generalized to other  applications  because it is in theory useful for any adaptation task which is facing the imbalanced data challenge between source and target datasets.

Our contributions can be summarized as follows:
\begin{itemize}
    \item We propose an unsupervised fault diagnosis framework which builds on simple fault generation process but performs well with the help of domain adaptation. The generation-adaptation framework makes use the power of both expert knowledge and domain adaptation to relieve the need of fault labels.
    \item A novel imbalance-robust domain adaptation approach is proposed for unsupervised fault diagnosis which is robust against different imbalance levels between different health conditions. 
    \item To our knowledge, we provide the first publicly-available synthetic dataset for bearing fault diagnosis. We also provide our code to facilitate further research. 
\end{itemize}

\section{Background}

\begin{figure*}
    	\centering
	{\includegraphics[width=1.65\columnwidth]{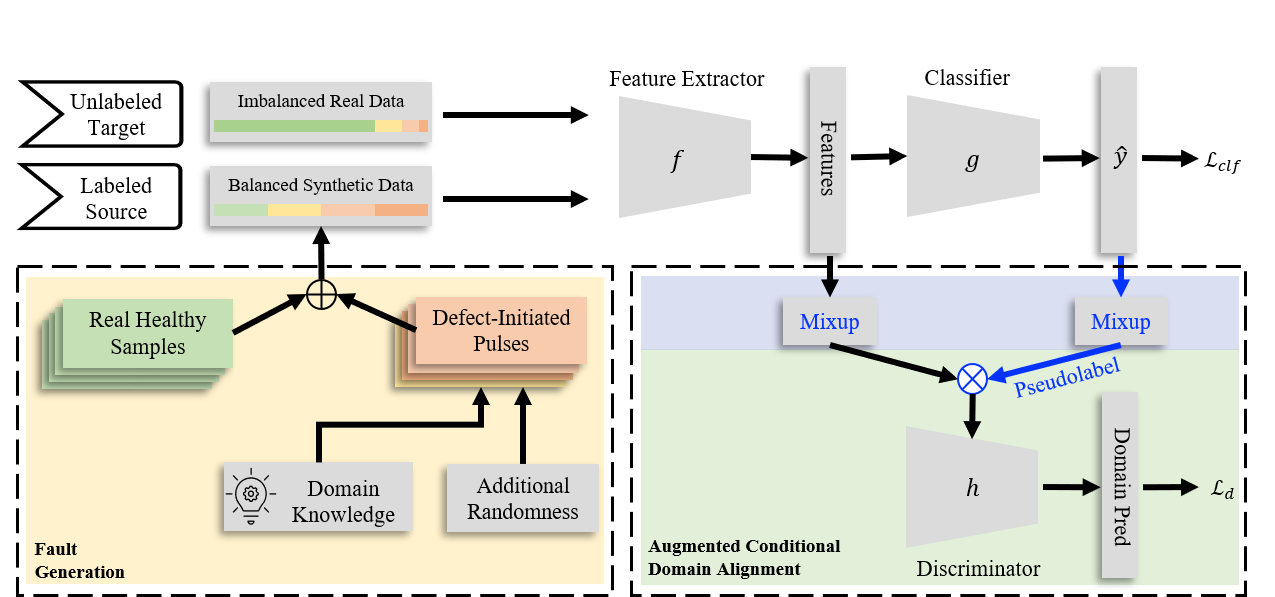}}
	\caption{Overview of the proposed method. (left) The fault generation method we used in stage one. (right) The proposed augmented conditional domain alignment method. As shown in the green block, both the feature and class information are fused by a multi-linear map before feeding into the discriminator to provide information on the class-conditioned distributions. In the standard adversarial approach DANN, the discriminator does not have access to class information $\hat{y}$. Our mixup-based distribution augmentation is shown in the purple block to provide additional distribution support for rare faults. The difference between our proposed augmented conditional domain alignment and DANN is shown in blue.}
	\vspace{-2mm}

	\label{fig2}
\end{figure*}
\subsection{The Use of Synthetic Data for Fault Diagnosis}
In the field of fault diagnosis, synthetic faults have been used as a reasonable substitute when data on real faults is not available{~\cite{eklund2006using}}. For example, synthetic faults can be induced from healthy samples using an analytical model~\cite{zhang2001wavelet}. Faults can also be simulated by highly accurate physical simulation such as finite element method (FEM) models{~\cite{gao2020fem}}.  In most works, the learning of the models is purely based on the synthetic faults, while the rare fault data collected from the real applications is preserved as a test set to evaluate the model performance~\cite{gryllias2012support, sobie2018simulation}. Some recent research studies, such as {\cite{liu2018small}} make the assumption that they have access to a very small set of labeled real faults. A large number of synthetic faults is then generated to mimic the real faults {by adversairal networks~\cite{ mao2019imbalanced} or interpolating between samples~\cite{zhang2018imbalanced}}. The corresponding data-driven models can be trained based on these imitated faults. This approach is different from previous works which learn solely from the synthetic data, because the model now also have an implicit or explicit access to information of the small real set of labeled faults. However, in these works, the unavoidable domain shift between the synthetic and real faults is still overlooked. In addition, the requirement of access to the labeled real faults imposes a real limitation on the generalization ability. If the target operating condition is different from the observed one, the learned method may fail because of the existence of a further gap between operating conditions. To the best of our knowledge, these domain gaps between synthetic and real data in the field of fault diagnosis has been long overlooked. Very recently, ~\cite{liu2020domain} used both synthetic data and real data to learn a model for remaining useful life prediction in the related field of prognostics. Our work contributes to this by considering a mix of both the healthy and synthetic data which will result in a more realistic source domain.

\subsection{Domain Adaptation}
Unsupervised domain adaptation{~\cite{pan2011domain}} is a powerful tool to alleviate the domain shift between synthetic and real data. DA approaches often consider the case where there is a labeled source domain~(synthetic in our case) and an unlabeled target~(real) domain. The methods improve the performance on the target by making use of both labeled source data and additional unlabeled target data. DA has been widely studied in fields such as computer vision{~\cite{ csurka2017domain}} and natural language processing{~\cite{jiang2007instance}}. The alleviation of the distribution difference between source and target is often the motivation for applying DA approaches. By aligning the distributions, models can effectively benefit from both the source and target data. Following this motivation, a series of approaches have explored different ways of alignments. Discrepancies such as Maximum Mean Discrepancy (MMD) were used~\cite{long2015learning} as a guide to align the distributions. {Distributions can also be aligned in normalization layers~\cite{li2016revisiting}.} Domain Adversarial Neural Networks (DANN)~\cite{ganin2014unsupervised} use a domain discriminator to adversarially align the features from source and target domains. The method usually have a classifier branch and a discriminator branch. Given a feature extractor $f$ parameterized by $\theta_f$, a classifier $g$ parameterized by $\theta_g$, and a discriminator $h$ parameterized by $\theta_h$, the DANN method is essentially solving the following equations: 
\begin{align}
\mathcal{L}(\theta_f, \theta_g, \theta_h) &= \mathcal{L}_{clf}(g(f(x))) - \lambda_d \mathcal{L}_d(h(f(x))),\label{dann0}\\
\hat\theta_f, \hat\theta_g &= \arg\min_{\theta_f, \theta_g} \mathcal{L}(\theta_f, \theta_g, \hat\theta_h),\\
\hat\theta_h &= \arg\max_{\theta_h} \mathcal{L}(\hat\theta_f, \hat\theta_g, \theta_h),
\end{align}
where $\mathcal{L}_{clf}$ is the cross entropy loss function for the main classifier. $\mathcal{L}_{d}$ is the cross entropy loss for the domain classification. The classifier branch is trained to minimize the classification loss $\mathcal{L}_{clf}$ on source data. The discriminator is trained to generate unbiased features for both source and target data. One the one hand, the optimization is updating the discriminator's weight by minimizing discriminator loss $\mathcal{L}_{d}$ to make it distinguish the source and target features as well as possible. On the other hand, the equation is forcing the feature extractor $f$ to generate unbiased features such that the discriminator loss is large. This minimax game effectively aligns the distributions between source and target. CDAN~\cite{long2018conditional} takes this concept one step further and conditions the adaptation models on discriminative information conveyed in the classifier predictions. 

{In recent years, input-space adaptation and self-training frameworks have also attracted many research interests. Aligning source and target data in the input space can be beneficial via Fourier transforms~\cite{yang2020fda}. In self-training frameworks, model performance can be gradually improved by iteratively training the network using target pseudo-labels as ground truth. Following this motivation, PyCDA~\cite{Lian_2019_ICCV} improves adaptation performance by training with pseudo-labels from different scales. Zhang et al.~\cite{zhang2021prototypical} proposes to refine pseudo-labels by making use of prototypes. Wang et al.~\cite{wang2021domain} find that explicitly learning the relationship between the main task and self-supervised auxiliary task can help to improve domain adaptation performance. Mei et al.~\cite{mei2020instance} find it beneficial for domain adaptation to integrate adversarial alignment with self-training.}

\subsection{Domain Adaptation for Fault Diagnosis}
In recent years, domain adaptation methods have raised a strong interest in the fault diagnosis community~\cite{zhang2017new, li2019multi , yang2019intelligent, 8643085, lu2016deep}. {Most existing works focus on the adaptation between operating conditions, and have found classic DA methods beneficial~\cite{wang2019domain}. Adversarial alignment~\cite{zhang2018adversarial} is widely used by exisiting works~\cite{wang2020intelligent, li2021domain}, and improved by conditional discriminators~\cite{zhang2021conditional, yu2020conditional}. Generative Adversarial Networks(GAN) are also explored to generate faults~\cite{li2018cross} to alleviate domain gap. Liu et al.~\cite{liu2021optimal} make use of optimal transport theory to align the distributions. Chen et al.~\cite{chen2020domain} propose to also take contribution of individual data samples into account, in addition to the global data distribution. We refer the readers to~\cite{yan2019knowledge} for an overview of existing methodologies and~\cite{zhao2019unsupervised} for implementation collections. More recent works focus on making the domain adaptation setup more realistic. For example, Zhu et al.~\cite{zhu2020new} focuse on learning from multiple source domains and Zhang et al.~\cite{zhang2021universal} consider the more general case of universal domain adaptation~\cite{you2019universal}.}
\\While many works make the assumption that both domains have a balanced number of samples for each class, \cite{wang2020missing} explores the scenario where the target data has missing fault classes. Another related line of work is to adapt models between different units of a fleet, particularly focusing on the complementary operating conditions~\cite{michau2021unsupervised}. Synthetic-to-real adaptation is relatively new and an unexplored research direction in the field of fault diagnostics. To the best of our knowledge, there has been no prior work which explores it for fault diagnosis where the imbalanced faults impose a great challenge to the domain adaptation task.

\section{Proposed Methodology}
To exploit the possibility of effectively learning and adapting from synthetic data in real industrial scenarios, we consider a challenging and  realistic setup where we only have access to segments of condition monitoring vibration recordings. Given unlabeled training data from the target real domain:  \begin{eqnarray}\mathcal{D}_{t}=\{x_{t}^1, ...,x_{t}^m\}, \end{eqnarray}

where $x_{t}^i$ is a vector of a real vibration recording. The target in this unsupervised fault diagnosis task is to train a model which performs well on this target domain.

\subsection{Overview of the Proposed Framework}
The proposed framework integrates expert knowledge with imbalance-robust domain adaptation for unsupervised fault diagnosis tasks. 

In the first part, the fault generation module makes use of expert knowledge on the fault types and generates synthetic faults based on the unlabeled real recordings of the bearings. Unlike previous works which utilize highly accurate physical models to synthesize the faults, our generation stage is deliberately designed to be simple such that the overall framework does not rely on highly accurate physical simulators. 

In the second part, a novel imbalance-robust domain adaption approach is proposed  to alleviate the distribution gap between the synthetic and real features, specifically for fault diagnosis. Unlike classic DA methods which typically deal with balanced data, the proposed method is able to align the features when the class distribution is different between the synthetic source and real target. This imbalance-robust framework is proposed for realisitic scenarios as real datasets  are often very imbalanced because of the rare faults. We would like to highlight that the proposed domain adaptation approach does not rely on specific generation process in the previous step. Thus, it can be easily generalized to other fault generation process or even to other classification problems that suffer from a class-imbalance between source and target datasets. 

Integrating both parts together leads to our proposed framework which makes effective use of both expert knowledge and unlabeled target data. The overview of our proposed framework is shown in Figure~\ref{fig2}.

\subsection{Bearing Fault Generation}
\label{section:stage1}

The challenge of fault generation in our proposed framework is how to make the synthetic faults adaptable to the given target~(real) domain without using highly accurate physical models. It is important to take the limited information provided in the unlabeled target samples into account. As a prerequisite, we assume that we have access to some real healthy samples. This is usually achievable by using the early recordings of each asset where we can safely assume that the faults have not yet emerged. Thus, this is a realistic assumption also in real application conditions. The healthy real samples are then used as the base signal for the synthetic process and inject the fault patterns using expert knowledge. Since the base signal encodes information about the operating and environmental conditions, the generated signals can be more adaptable to the target domain.

To this end, a general procedure is followed to generate synthetic bearing defect signals as described in \cite{randall2001}. Let $s$ denote the oscillating waveform excited by a single impact due to over-rolling a surface defect with period $T$. The amplitude of the $i^{th}$ impact is denoted by $A_i$ with period $Q$, which mimics a modulation term due to a rotating inner ring or a rolling element defect. The terms $T$ and $Q$ can be directly calculated based on the kinematics of the bearing and a speed recording by a domain expert (see Eqs. 1-4 in \cite{randall2011}). We can now describe our modeled vibration signal at time $t$ by,

\begin{equation}
\epsilon(t) = \sum_{i=-\infty}^{\infty}{A_i s(t-iT) + \beta n(t)},
\end{equation}

\noindent where the additive background noise term $n$ is taken equal to a vibration recording of a healthy bearing. To make our method robust for various signal-to-noise ratios (SNRs), $\beta$ is uniformly distributed between 0.25 and 2. A single impact $s$ is modeled by means of a Hann-window with a duty period of 5\% with respect to $T$. Since we do not know the transfer function, a wide-band band-pass filter is applied to $s$ instead. This filter mainly attenuates frequencies close to 0 Hz and the Nyquist frequency. As a result, our fault frequencies are visible in any frequency band, which facilitates the domain adaptation to bridge the gap with the real sensor data. The periodic amplitude modulation term $A_i$ is expressed as a sum of cosines,

\begin{equation}
A_i = \gamma \sum_{k=0}^{K}{\alpha_k \cos{(iT k2\pi/Q)}},
\end{equation}

\noindent such that we can control the number of side-bands $k$ with corresponding amplitudes $\alpha$ in our envelope analysis directly. To introduce some natural randomness of the impacts, $\gamma$ is normally distributed with unit mean and a standard deviation of 0.1. In our work we use $\alpha=[1, 0.76, 0.38, 0.11, 0.05]$. The chosen $\alpha$ should describe a typical decaying side-band pattern for a radially loaded bearing. We would like to emphasize that $\alpha$ was not specifically optimized to improve the diagnosis performance on any of our datasets as we don't assume access to fault labels in the real data for training. Hence, a domain expert does not need exact knowledge of $\alpha$. However, in case of an axially loaded bearing one could consider reducing the number of side-bands (see~ \cite{mcfadden1984} for more details). The proposed generation method follows similar procedures as introduced in~\cite{randall2001,borghesani2013}.

\subsection{Augmented Conditional Adversarial Alignment}
The proposed generation approach provide us with the source dataset $\mathcal{D}_s$ which contains the synthetic faults. To make efficient use of the unlabeled data from the real target domain, we propose to re-formulated the unsupervised fault diagnosis task as a unsupervised domain adaptation problem. 
\begin{itemize}
    \item Synthetic source domain data with balanced samples across all classes
    \begin{eqnarray}\mathcal{D}_s=\{(x_{s}^1, y_{s}^1), ...,(x_{s}^n, y_{s}^n)\},  y_{s}^i \in Y.\end{eqnarray}
    \item Unlabeled real data from target domain with an imbalance across classes  \begin{eqnarray}\mathcal{D}_{t}=\{x_{t}^1, ...,x_{t}^m\},\end{eqnarray}
    \end{itemize}
where $Y$ is the set of discrete health states. Our aim is to improve the model performance in the target domain $\mathcal{D}_{t}$. The setup is now similar to transductive DA problems~\cite{gammerman1998learning, arnold2007comparative}. 

\subsubsection{Imbalance Issue for Direct Alignment}
There is one essential difference between the standard DA commonly seen in image classification and our setup which could potentially harm the alignment quality. In most DA setups, the source and target domains are assumed to have the same class distributions, meaning that either both source and target are class balanced, or both follow a similar class distribution. However, for our synthetic-to-real setup, the scenario is quite different. On the synthetic data side, since faults are all generated based on the healthy samples, each health condition has the exact same number of samples, leading to a balanced source dataset. On the target real data side, the class distribution contains a majority of healthy states and few faults, leading to an imbalanced dataset. 

\begin{figure}
    	\centering
	{\includegraphics[width=0.9\columnwidth]{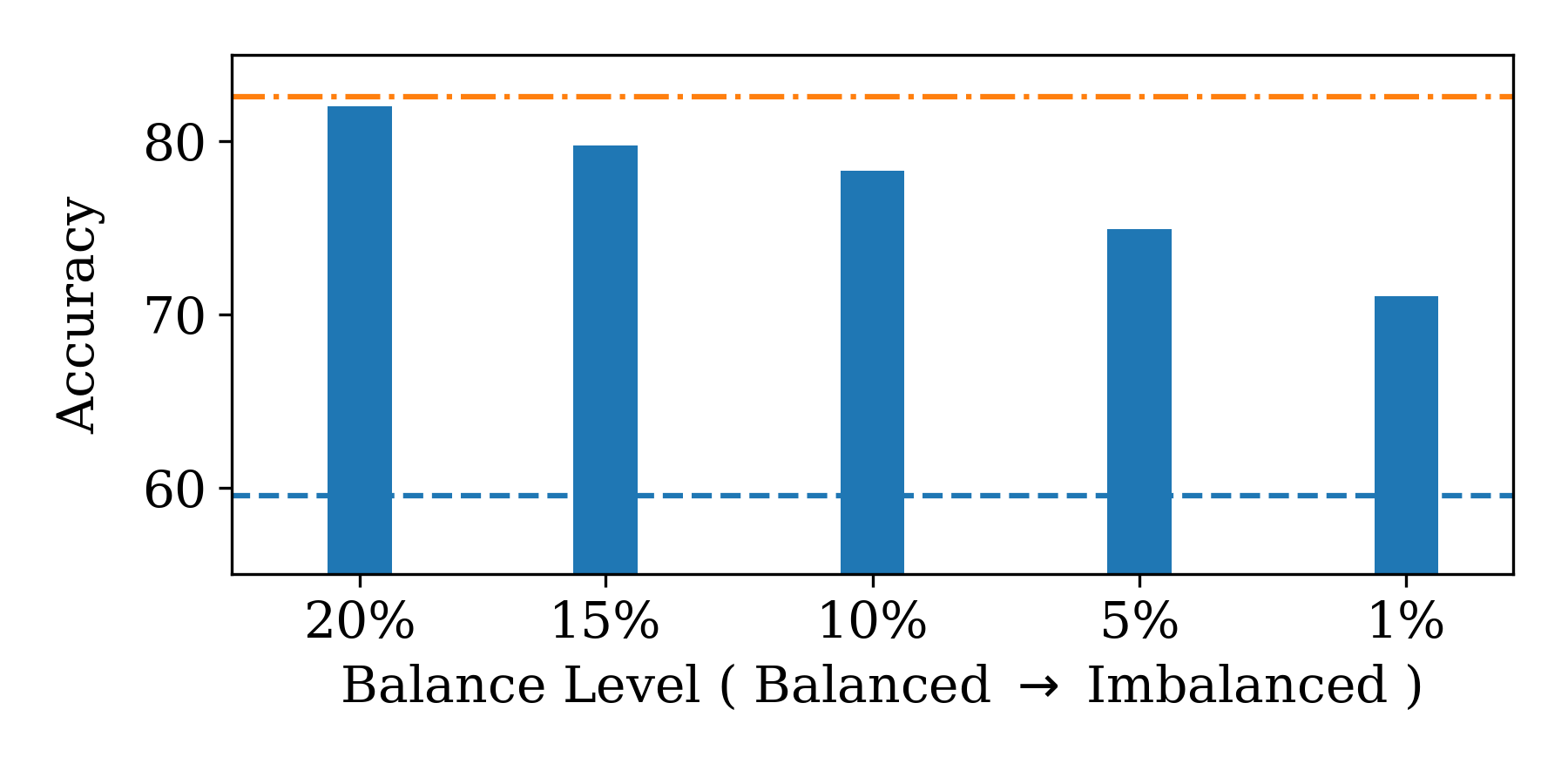}}
	
	\vspace{-5mm}
	\caption{Standard adversarial domain adaptation performance decreases more rapidly when the target domain becomes more imbalanced. A synthetic-to-real CWRU experiment with different levels of imbalance on the rolling element fault class. 1\% on the x-axis means that only 1\% samples are available for rare fault class, compared to the balanced case.  The orange dash-dot line indicates the performance when the dataset is fully balanced~(100\%). Blue dash line indicates the source-only baseline. }
	\label{fig1}
	\vspace{-5mm}
\end{figure}

This mismatch between the source and target domain can in fact lead to a significant decline of the adaptation performance. We show empirically in Figure~\ref{fig1} the performance of a naive synthetic-to-real adversarial adaptation based on the CWRU dataset using DANN~\cite{ganin2014unsupervised}. In this setup, the source synthetic domain is balanced, while for the target domain, one of the faults~(rolling element fault) has a smaller number of samples. The experiment shows that, compared to a fully balanced case, the stronger class imbalance in the target leads to a steeper performance decline.

To answer why such a performance decline occurs, we need to look back into the assumptions that the classical DA methods make. Classical DANN methods~\cite{ganin2014unsupervised, ganin2016domain, wang2019domain} are based on the following domain adaptation theory: the error function of the domain discriminator corresponds well to the discrepancy between unconditional feature distributions $P$ and $Q$. Thus, minimizing the error rate of the discriminator can lead to aligned feature distributions. However, when the joint distributions of feature and class, i.e. $P(x_s, y_s)$ and $Q(x_t, y_t)$, are non-identical across domains, adapting only the feature representation may be insufficient~\cite{long2018conditional}. This can be an especially large problem when the target class distributions in fault diagnosis are heavily skewed towards the healthy class. In this case, simply aligning the features, in fact, gives no theoretical guarantee that two different distributions are identical even if the discriminator is fully confused~\cite{arora2017generalization}.

\subsubsection{Class Information Taken into Account}
The performance degradation stems from the fact that the class distributions are mismatched and the discriminator fails to take any class information into account. We, thus, propose to use the class information of the representations as additional input to the discriminator in order to counteract the class mismatch from rare faults. By providing the generator hints on the class information, the discriminator can better align the distributions.

We use the multi-linear map to fuse the class information with features to improve the domain adaptation performance under severe class imbalance. As shown in Fig~\ref{fig2}, given a feature extractor $f$ parameterized by $\theta_f$, a classifier $g$ parameterized by $\theta_g$, and a discriminator $h$ parameterized by $\theta_h$, the loss function then becomes:
\begin{eqnarray}
\mathcal{L}(\theta_f, \theta_g, \theta_h) = \mathcal{L}_{clf}(g(f(x))) - \lambda_d \mathcal{L}_d(h(f(x)\otimes g(f(x)))),\label{eqmain}
\end{eqnarray}

where $\otimes$ is a multi-linear map, and $\hat{y} = g(f(x))$ represents the predicted pseudo-labels. Comparing to equation~\ref{dann0} of the standard adversarial training, the main difference is that instead of using features $f(x)$ alone as input to the discriminator,  the information of features and classes~(provided by pseudo-labels) are combined together via a multi-linear map. By applying the map, the discriminator can gain the information from both the feature distribution and class distribution and better align the class conditioned distributions. 

The proposed method is inspired by conditional adversarial methods in other application fields. For example, for image generation tasks, Conditional Generative Adversarial Networks~\cite{mirza2014conditional} concatenate the class vector with the feature vector to generate images conditioned on the single classes. In domain adaptation for image classification, CDAN~\cite{long2018conditional} uses multi-linear conditioning to align multi-modal distributions. These methods often deal with balanced datasets and does not focus on severe class-imbalanced scenarios, but the idea of aligning the conditional distribution is especially suitable for fault diagnosis.

\subsubsection{Augmented Distributions}
One potential issue with the above solution is that even though the important class information is provided for the discriminator, the fact that the rare fault classes have so few samples can potentially make the optimization of the discriminator unstable. This can also lead to a decline of the alignment performance.  To reinforce the alignment, it is essential to provide better distribution support for the conditional distributions on the rare classes. We propose to augment the features and pseudo-labels used as input to the discriminator. In particular, for a batch of $x$ from the target real dataset, we can get its corresponding feature embedding $e=f(x)$, and pseudo-label $\hat{y}=g(f(x))$ vectors. Whereby, $\hat{y}$ is generated by using the prediction from the model trained in the previous iteration. Then, the augmented interpolated sample can be represented as,
\begin{eqnarray}
\tilde{e} &=& \lambda e + (1-\lambda) e[\text{idx}],  \label{eqn:interp_x}\\
\tilde{y} &=& \lambda \hat{y} + (1-\lambda) \hat{y}[\text{idx}],\label{eqn:interp_y}\\
\tilde{z} &=& \tilde{e} \otimes \tilde{y},
\end{eqnarray}
where $\text{idx}$ is the shuffled index of a batch, $\lambda$ is generated from an prior $\beta$ distribution, i.e. $ \lambda \sim \beta(\alpha, \alpha)$  with $\alpha$ controls the shape of the $\beta$ distribution, $\tilde{e}$ is the interpolated features, $\tilde{y}$ is its class label, and $\tilde{z}$ is the multi-linear input to the discriminator. We use $\alpha=1$ for all our experiments. This idea is inspired by MixUp~\cite{zhang2017mixup} in image classification, where the authors mix input images and labels to provide more training samples in a fully supervised training setup. Instead of directly mixing the input images in MixUp, our proposed distribution augmentation is conducted on the feature space and used for our unsupervised domain adaptation setup.

The motivation behind this is to augment the conditional feature distributions of rare classes for fault diagnosis. By mixing up the features within a batch, the fault information becomes present in more samples within a batch. The interpolated samples enlarge the target training dataset for the rare fault classes, making the learning process for the discriminator more stable. We highlight the difference between DANN and our alignment method in blue on the right side of Figure~\ref{fig2}.

\section{Datasets}
To facilitate our adaptation experiments, as described in Section~\ref{section:stage1}, we generate the synthetic faults for two datasets.

\begin{figure}
\centering
\includegraphics[trim=0.4cm 0cm 0.4cm 0, clip=true, width=1\columnwidth]{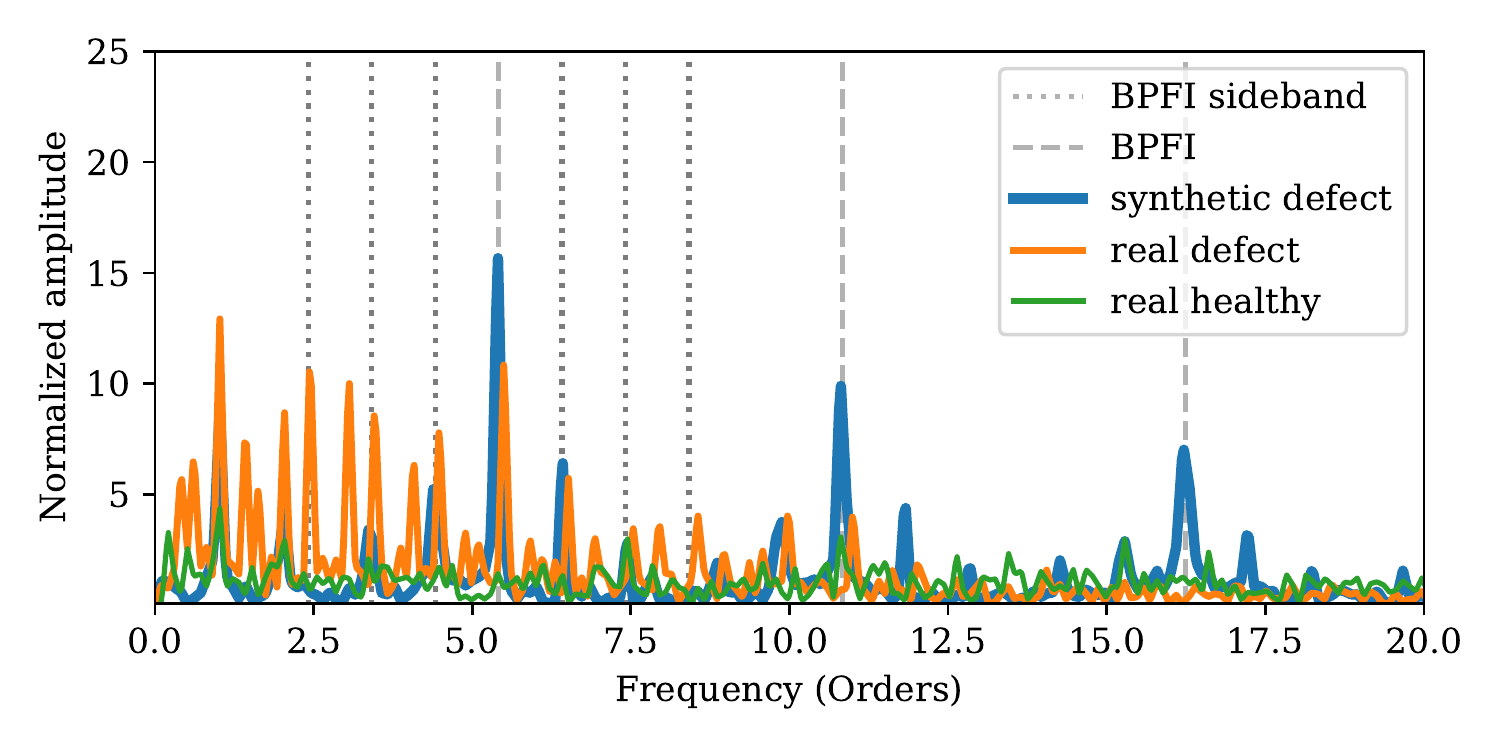}
	\vspace{-10mm}
\caption{CWRU normalized {full-wave rectified envelope spectrum} of an inner ring defect example with corresponding defect ball pass frequency (BPFI) denoted by the dashed vertical lines. The difference between the real and synthetic fault in BPFI frequencies and sidebands (showed by vertical dotted lines) motivates the use of domain adaptation.}
\label{cwru:example}
	\vspace{-4mm}
\end{figure}

\begin{figure}
\centering
\includegraphics[trim=0.4cm 0cm 0.4cm 0, clip=true, width=1\columnwidth]{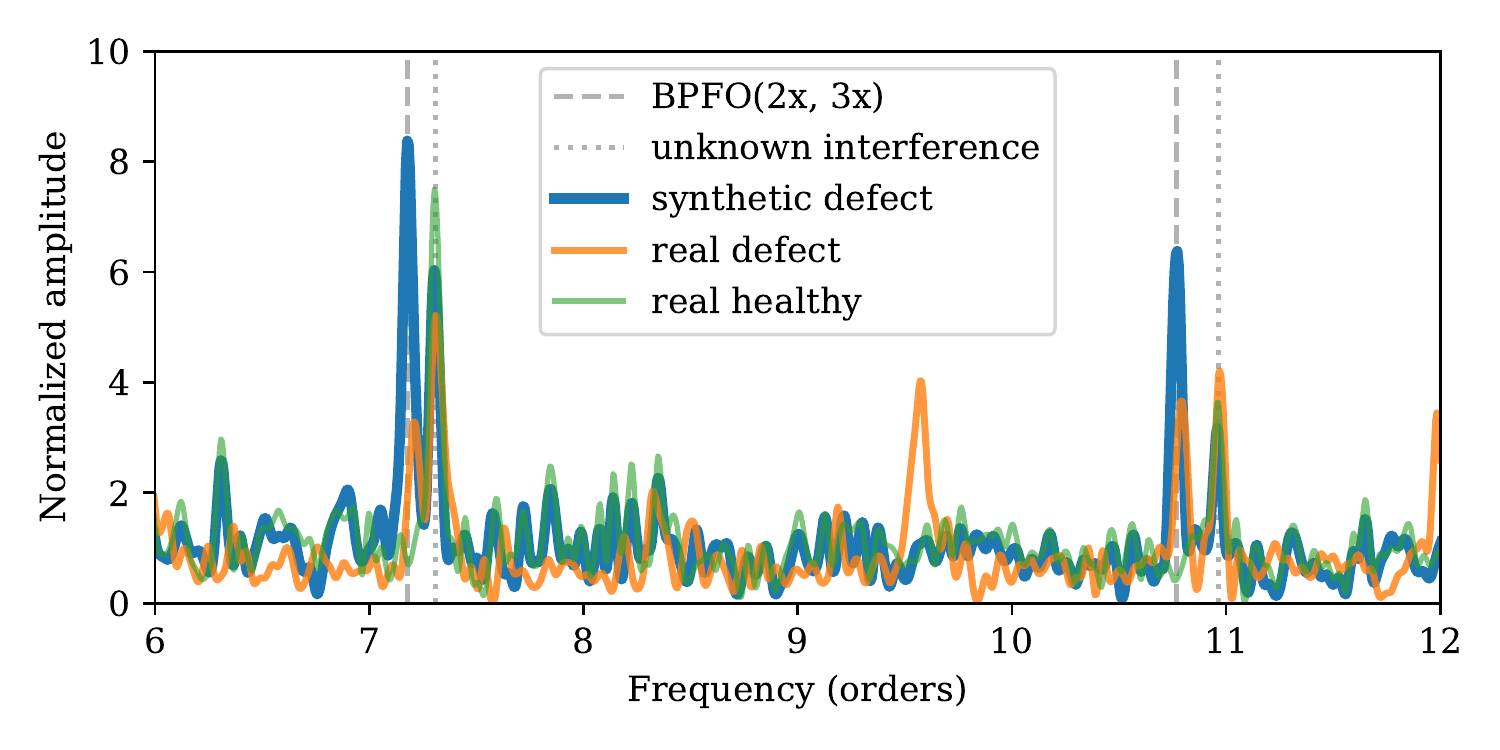}

	\vspace{-5mm}
\caption{Wind turbine generator dataset: normalized {full-wave rectified envelope spectrum}  of an outer ring defect example. In this example, the unknown interference is realistically transferred from healthy to synthetic. However, the synthetic fault does not match perfectly with the real fault, and the difference motivates the use of domain adaptation. Notice the frequency range between 6-12 orders such that only two and three times the ball-pass frequency of the outer ring (BPFO) is visible. }
	\vspace{-5mm}

\label{turbine:example}
\end{figure}
\subsection{CWRU Bearing Data}
The Case Western Reserve University (CWRU)~\cite{smith2015rolling} bearing dataset is used. It is a benchmark dataset for DA in bearing fault diagnosis~\cite{lu2016deep, zhang2018adversarial, wang2019domain}. Drive-end accelerometer data of 12 kHz sampling rate is used. We consider four health states~(classes) in this paper: healthy, inner race fault (IF), rolling element fault (REF), and outer race fault (OF). We group the sub-fault-types of different spall sizes together. For each health state, we sample 1200 segments. Each signal segment contains 4096 points. This results in 4800 samples for the real dataset. Subsequently, our generation method is applied to the real healthy signals to synthesize the real defects. To avoid data leakage during evaluation, we split half of the real healthy samples to the synthetic source data and up-sample the number of samples in healthy class in both domains back to 1200.

Figure~\ref{cwru:example} shows the {full-wave rectified envelope spectrum} of one synthetic and one real example of the inner race fault. Defect frequencies are indicated by the dashed lines, where a dotted line shows the first three side-bands of an inner ring defect around its ball-pass frequency (BPFI). The BPFI and its first harmonic in the synthetic and real examples are roughly aligned. However, differences can still be observed: 1) the real defect has more side-bands around the BPFI 2) the 2\textsuperscript{nd} harmonic is not present in the real defect and 3) the defect frequencies have a slightly higher frequency compared to the analytic defect frequencies. These differences motivate the use of domain adaptation to bridge the gap between the real and synthetic signals.

\subsection{Wind turbine bearing data}
A real-world dataset from generator bearings of multiple wind turbines is collected and used to evaluate the method. 
This data origins from a condition monitoring service, where bearings are monitored by human experts aided by analysis software. For each turbine, historic data is available varying between 3-5 years, where on a daily basis two recordings are made. Each vibration recording has a length of 1.28 seconds and a sample-rate of 12.8 kHz. Rotational speeds vary between 900 and 1700 RPM. Speed recordings are available via a tachometer. There are three health states~(classes),  healthy, outer-ring defects, and inner-ring defects.

Based on a combination of manual analysis of vibration and process data, knowledge of bearing replacements and anonymized customer feedback on bearing defects, the data was carefully labeled. In total, we collected 1643 samples for the healthy state, 2990 samples for the outer-ring defects, and 192 samples for the inner-ring defects from the generator drive-end bearings each on a different wind turbine. Note that during the labeling of the data we had access to additional process data such as electrical power and wind-speed but were also able to inspect trends and high SNR averaged spectra of many vibration recordings. As a consequence, a single vibration recording, as used by our proposed domain adaptation method, is only a fraction of the information used in the labeling procedure. In many situations, all these additional information sources are unavailable, which motivates the use of single vibration recording predictions as in this research.

\section{Evaluation}
\subsection{Implementation}
\subsubsection{Model Architecture}

\begin{table}[]
\centering
\caption{Network components used in all our experiments. }
\vspace{-2mm}
\begin{tabular}{ccccc}
\hline
\multicolumn{5}{c}{\textbf{Input $x$}}                                                                                                         \\ \hline\hline
\multicolumn{5}{c}{\textbf{Feature Extractor $f$}}                                                                                             \\ \hline
\multicolumn{1}{c|}{Conv0} & \multicolumn{4}{c}{filter size 3; channel num 10}                                                             \\
\multicolumn{1}{c|}{}      & \multicolumn{4}{c}{relu; dropout}                                                                             \\ \hline
\multicolumn{1}{c|}{Conv1} & \multicolumn{4}{c}{filter size 3; channel num 10}                                                             \\
\multicolumn{1}{c|}{}      & \multicolumn{4}{c}{relu; dropout}                                                                             \\ \hline
\multicolumn{1}{c|}{Conv2} & \multicolumn{4}{c}{filter size 3; channel num 10}                                                             \\
\multicolumn{1}{c|}{}      & \multicolumn{4}{c}{relu; dropout; flatten}                                                                    \\ \hline
\multicolumn{1}{c|}{Dense} & \multicolumn{4}{c}{channel num 256}                                                                           \\
\multicolumn{1}{c|}{}      & \multicolumn{4}{c}{relu; dropout}                                                                             \\ \hline
\multicolumn{5}{c}{\textbf{256-D Features $f(x)$}}                                                                                                \\ \hline\hline
\multicolumn{2}{c|}{\textbf{Classifier $g$}}                          & \multicolumn{1}{l|}{} & \multicolumn{2}{c}{\textbf{Discriminator $h$}}     \\ \cline{1-2} \cline{4-5} 
\multicolumn{1}{c|}{Dense} & \multicolumn{1}{c|}{channel num 256} & \multicolumn{1}{c|}{} & \multicolumn{1}{c|}{Dense}  & channel num 512  \\
\multicolumn{1}{c|}{}      & \multicolumn{1}{c|}{relu}            & \multicolumn{1}{c|}{} & \multicolumn{1}{c|}{}       & relu             \\ \cline{1-2} \cline{4-5} 
\multicolumn{1}{c|}{Dense} & \multicolumn{1}{c|}{num of classes}  & \multicolumn{1}{c|}{} & \multicolumn{1}{c|}{Dense}  & channel num 2    \\
\multicolumn{1}{c|}{}      & \multicolumn{1}{c|}{softmax}         & \multicolumn{1}{c|}{} & \multicolumn{1}{c|}{}       & softmax          \\ \cline{1-2} \cline{4-5} 
\multicolumn{2}{c|}{\textbf{Class Prediction}}                    & \multicolumn{1}{l|}{} & \multicolumn{2}{c}{\textbf{Domain Prediction}} \\ \hline
\end{tabular}
	
\label{arch}

	\vspace{-5mm}
\end{table}

The network consists of a feature extractor, classifier, and a discriminator. The architecture of the feature extractor and classifier are taken from~\cite{li2018cross}. The details are summarized in Table~\ref{arch}.

\subsubsection{Data Preprocessing}
A preprocessing step is applied to all data. 
First, the time-domain waveforms are normalized to unit standard deviation and subsequently converted to an {full-wave rectified envelope spectrum}. It is obtained by taking the Fourier transform magnitudes of the full-wave rectified and band-pass filtered signals. {A band-pass filter with a pass-band between 500-4000 Hz is used following a standard frequency band for bearing diagnostics as described in \cite{morris2016}. We filter at 4000 Hz to attenuate the signal below the Nyquist frequency. Although more optimal bearing diagnostic signal representations are available, e.g., as used in \cite{chen2020}, this simple representation is sufficient to make a valid comparison between the different domain adaptation methods. As no labeled real-world faulty signal is available, Kurtogram based methods~\cite{antoni2016, wang2016spectral} cannot be used to optimize the pass-bands.} Subsequently, the {full-wave rectified envelope spectrum} is then interpolated to a speed-normalized axis of 1000 values between 0-30 repetition orders. {The same data preprocessing is applied on all experiments across all methods.}

\subsubsection{Evaluation Metric}
Since the classes are imbalanced, we report a balanced version of the accuracy. This metric was used by imbalanced classification tasks such as~\cite{li2017webvision}. For a dataset with $K$ classes, the reported accuracy is defined as: 
$$\frac{1}{K} \sum_{k=1}^{K}{\frac{P_k}{M_k}} $$
where $M_k$ is the number of test samples with class $k$ as ground truth, and $P_k$ is the number of correct predictions for the given $M_k$ samples in class $k$. This metric provides a fair comparison also in highly imbalanced datasets. All reported results are based on the average of ten runs  using this metric. {We additionally report the F1 score and Cohen's kappa~\cite{cohen1960coefficient}. We use the implementation from Scikit-learn~\cite{scikit-learn} for all evaluation metrics.}

\subsubsection{{Training and Test Details}}
\begin{figure}
    	\centering
	{\includegraphics[width=0.85\columnwidth]{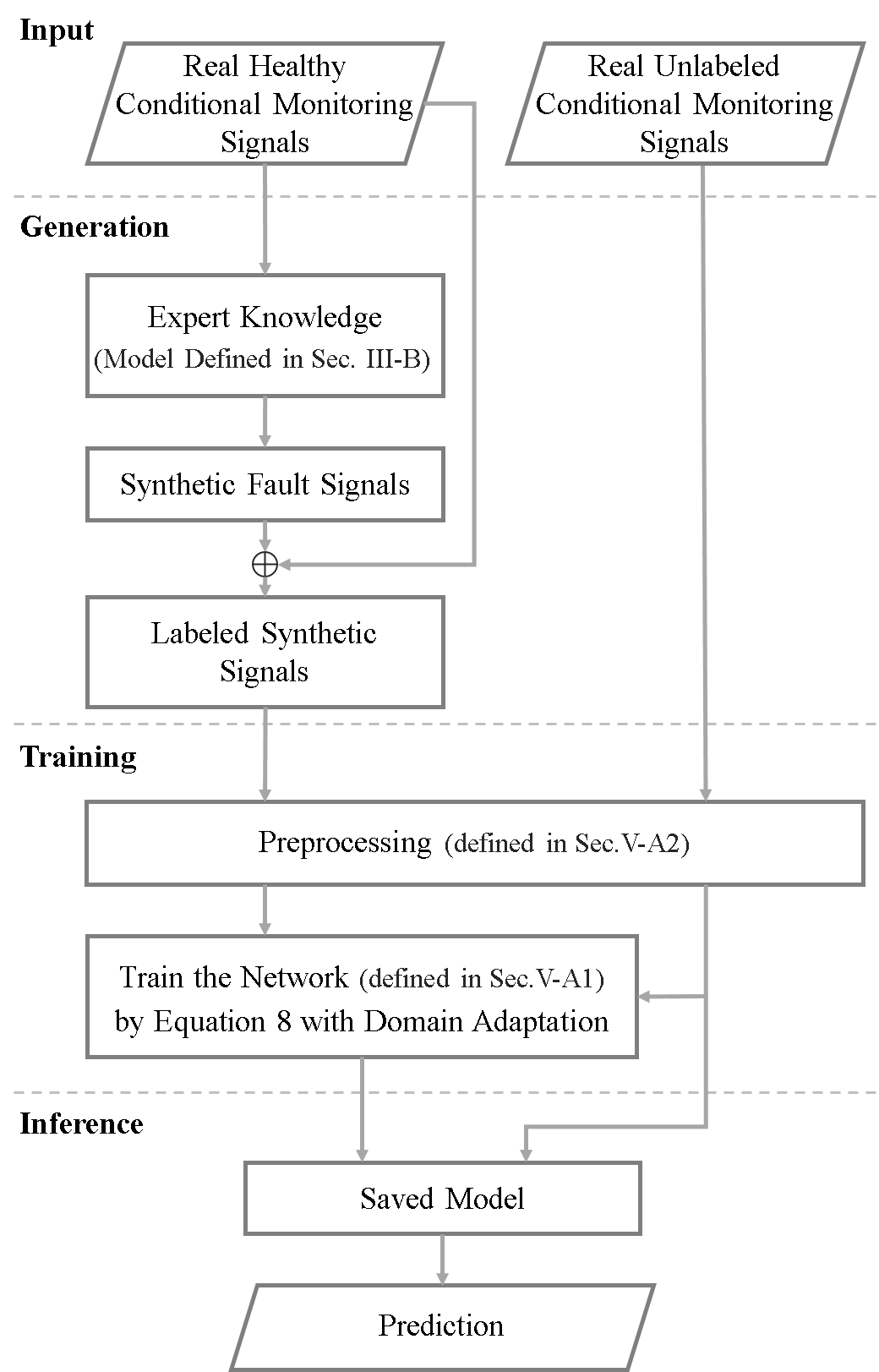}}
	\caption{{The flow chart of the proposed framework.}}
	\label{flowchart}
	\vspace{-1mm}
\end{figure}

{
We train our neural architecture specified in Table~\ref{arch} end-to-end for 100 epochs with a batch size of 128 and a learning rate of $0.001$. All models are trained using the Adam optimizer. The same architecture and training parameters are used for all methods in our experiments to ensure a fair comparison. During training, only the source labeled data $\mathcal{D}_s$ is used to update the classifier $g$ via $\mathcal{L}_{clf}$ because we don't have labels for the target real data $\mathcal{D}_t$. Both the synthetic and real data contribute to the learning of the feature extractor $f$ and Discriminator $h$ via the combined loss defined in Equation~\ref{eqmain}.  For inference, we feed the target data to the feature extractor $f$ and classifier $g$ to make the prediction. The discriminator is not needed in test time.  All models are trained and tested under different random seeds for 10 times and we report the average performance. A flowchart of the proposed framework is provided in Figure~\ref{flowchart}.}

\subsection{CWRU Synthetic-to-Real Experiments with Different Levels of Imbalance on One Fault.}
\setlength{\tabcolsep}{3pt}
\begin{table}[]

\begin{center}
\caption{Synthetic-to-real CWRU adaptation with different balance levels on rolling element faults. }

\label{ablation}
\begin{tabularx}{\columnwidth}{c|c|c|c|c}
\hline
Balance Level$^*$   &  DANN~\cite{ganin2016domain}    & Conditional(ours)  & Proposed(ours)& $\Delta$ \\ \hline
20\%                    & 82.04\% & 82.11\%    & 83.49\% &  +1.44\%\\ \hline
15\%                    & 79.77\% & 81.90\%    & 83.04\% &  +3.27\%\\ \hline
10\%                    & 78.30\% & 82.88\%    & 83.32\% &  +5.02\%\\ \hline
5\%                         & 74.96\% & 76.90\% & 83.30\%  & +8.34\%  \\ \hline
1\%                          & 71.05\% & 73.93\%  &  82.34\% & +11.29\%  \\ \hline
\end{tabularx}
\begin{tabularx}{\columnwidth}{Y|Y|Y|Y}
\hline\hline
 Source-only                  & 59.55\%    & Balanced DANN & 82.59\%%
 \\ \hline
  {Balanced CDAN} & {80.32\%} & {Balanced Ours} & {84.18\%}\\ \hline

\end{tabularx}
\end{center}
\footnotesize{$^*$ Smaller balance level indicates fewer samples and larger imbalance.}\\

	\vspace{-5mm}
\end{table}
We evaluate the proposed method against a source-only baseline and the standard DANN alignment. In this ablation study we reduce the complexity by assuming that all classes have the same number of samples except the rolling element fault. More realistic setups are evaluated in later sections. To show the effectiveness of the proposed methodology against different levels of class imbalance, we change the number of samples with the rolling element fault. For example, when the balance level is 1\%, the unlabeled target training data contains $1\% * 1200 = 12$ samples for the rolling element fault class, while 1200 samples are available for every other class. In the $10\%$ setup, the target set contains $10\% * 1200 = 120$ samples for the rolling element fault class. A smaller value of level of balance~(1\%) indicates a larger level of class imbalance as the number of faults is considerably smaller compared to the number of healthy class samples.  

As shown in Table~\ref{ablation}, without any alignment, the source-only accuracy is 59.55\% in this 4-class-classification problem. This is already higher than the random guess performance of 25\% and proves that our synthetic faults encode the fault information in a meaningful way. On the other hand, DANN can perform quite well when all the classes are balanced, but degrades dramatically when the level of balance decreases. By taking the class information into account, conditional alignment alone shows resistance to the degradation from the  class imbalance. However, the performance gap still exists, and the performance becomes less stable when the balance level decreases. By further augmenting the conditional distributions, the proposed method can significantly strengthen the quality of the alignment and lead to a performance that is at a similar level as for the balanced dataset, even when there are only $1\%$ rolling element fault samples in the target dataset. This demonstrates the effectiveness of our proposed method. 

We also report the improvement of the proposed method over the standard DANN. An interesting observation is that with an increasing level of imbalance, our method provides an increasing amount of improvement over DANN. This suggests that the method can be especially beneficial when imbalance in the target dataset is severe. Furthermore, the proposed method appears to be robust to the degree of imbalance. The results for all the degrees of imbalance demonstrate a similar accuracy. Even with a relatively strong imbalance ($1\%$ balance level) , the performance is similar to the the fully balanced setup of the standard DANN. Therefore, the methodology is suitable for cases where the degree of class imbalance in the target dataset is not known which is a typical setup in real applications. Thus, the method is quite suitable for real fault diagnosis tasks.

\paragraph{{Performance under a balanced setting}} {To further analyze the effectiveness of the proposed method, we report the performance under the balanced CWRU setup in Table~\ref{ablation}. Under this setup, all classes in the target domain have equal number of samples. We observe that all alignment methods work well comparing to the source-only baseline. The DANN methods achieves 82.59\%, while our method provides  further improvement and achieves 84.18\%.  This shows that the proposed augmented conditional alignment method is complementary to the adversarial alignment, even when all classes are balanced. This is most likely due to our feature distribution augmentation strategy which better facilitates the adversarial alignment.}

\subsection{CWRU Synthetic-to-Real Experiment with a More Realistic Class Imbalance among Faults}
\label{cc}
\begin{table}[]
\begin{center}
\caption{Experiment results on the synthetic-to-real task for CWRU dataset when level of balance is different among faults. The target data contains 1200~(100\%) healthy samples, 120~(10\%) OF samples, 60~(5\%) IF samples and 12~(1\%) REF samples. }
\begin{tabular}{c|c|c|c|c|c}
\hline
Metric & Source &  DANN~\cite{ganin2016domain}   & {IAST~\cite{mei2020instance}}& Condi. & Proposed \\ \hline
Balanced acc. & 0.596  & 0.726& {0.760} & 0.754  &\textbf{0.823}      \\\hline
{F1 (Macro)} & {0.402}  & {0.599} & {0.681}& {0.678}   & {\textbf{0.814}}     \\\hline
{F1 (Micro)} & {0.488}  &  {0.630} & {0.697}& {0.688}  & {\textbf{0.787}}     \\\hline
{Cohen's kappa} & {0.352}  & {0.476} & {0.574} & {0.560 } & {\textbf{0.696}} \\\hline
\end{tabular}
\label{cwru}
\end{center}\vspace{-0.4cm}
\end{table}

In reality, the imbalance does not only exist for one single fault, but also exists between different faults. To mimic also the imbalance between several faults in the target domain, we consider the following setup. The unlabeled target dataset consists of  $N=1200$ healthy samples, 10\% $\times$ N samples for the OF, 5\% $\times$ N samples for the IF and 1\% $\times$ N sample for the REF. This constructs a highly imbalanced target dataset. All labels are again removed during the training. This setup makes it a more challenging task as the imbalance levels of the three faults are different. 

Results in Table~\ref{cwru} show that when class imbalance is severe among different faults, the proposed method can still provide an improved alignment for the different classes. Compared to the simple DANN~\cite{ganin2016domain} alignment method which does not consider the imbalance between different classes, both the conditional alignment method CDAN~\cite{long2018conditional} and our proposed mixup augmentation provide a significant improvement. Conditional alignment alone provides a 2.8\% absolute improvement in balanced accuracy over DANN by simply providing additional class information. {This improvement is also validated by the F1-score and Cohen's kappa.} By combining the proposed augmentation with the class conditional alignment, we achieve a much stronger balanced accuracy of 82.3\% on the target data. This is a 9.7\% absolute improvement over the DANN baseline, and the improvement is validated also by the other metrics.  The overall performance of the proposed methodology is similar to that of DANN in the ideal fully balanced case. This makes the proposed framework applicable for unknown imbalance levels achieving same level of performance as in a balanced dataset. This finding is particularly encouraging for real applications where the imbalance level is unknown apriori. 

{We additionally compare our method with the state-of-the-art domain adaptation method IAST~\cite{mei2020instance}. IAST constitutes a stronger competing methods as it combines the advantages of adversarial learning and self-training. Under this setup with severe imbalance, our proposed  method outperforms IAST by a large margin in terms of all metrics including balanced accuracy, F1-score, and Cohen's kappa. }

\begin{table}[]
\centering
\caption{Experiment results  on the Wind Turbine dataset.}
\begin{tabular}{c|c|c|c|c|c}
\hline
 Metric & Source &  DANN~\cite{ganin2016domain}     &{IAST~\cite{mei2020instance}} & Condi. & Proposed \\ \hline
Balanced acc. & 0.609  & 0.645 & {0.679}& 0.683   &\textbf{0.708}      \\\hline
{F1 (Macro)} & {0.553}  & {0.570} & {0.666} & {0.671}  & {\textbf{0.683}} \\ \hline
{F1 (Micro)} & {0.681}  & {0.768} & {0.842}&{0.840} & {\textbf{0.845}}\\ \hline
{Cohen's kappa} & {0.411} & {0.524} &{0.643}& {0.641} & {\textbf{0.650}} \\\hline
\end{tabular}

	\vspace{-2mm}
\label{skf}
\end{table}
\subsection{Wind Turbine Generator Synthetic-to-Real Experiment}
To evaluate the effectiveness of the proposed methodology in a more realistic scenario, we additionally conduct an experiment on the real-world wind turbine bearings. As described in the dataset section, the outer race fault has a significantly larger number of samples than the inner race fault. In this setup, the source-only baseline achieves an accuracy of 60.85\%, showing that the synthetic data we generated for the wind turbine bearing is meaningful. The DANN method can improve the performance and achieve 64.5\%. By providing class information to the discriminator, we can achieve 68.3\%. If we additionally use our augmentation to enhance the distribution support, the final proposed method can achieve an accuracy of 70.8\%, yielding an almost 10\% absolute improvement on this real dataset, compared to the source-only baseline. {In terms of F1-score and Cohen-s Kappa, the proposed method also shows improvement over the source only baselind and DANN approach. The performance improvement is large over the baselines, but less significant than the setup in Section~\ref{cc}. This is mostly likely because the class imbalance is less severe in the wind turbine dataset than in the setup we used in Section~\ref{cc}. We additionally compare our method with the state-of-the-art domain adaptation method IAST~\cite{mei2020instance}. The proposed method is able to outperform IAST in terms of all four metrics. This validates the effectiveness and competitiveness of our proposed augmented conditional distribution alignment. }

We would like to emphasize that in all experiments, we only use knowledge of the healthy labels from the target domain. All class information on the faults is learned from the synthetic data where the expert knowledge is encoded. This result shows that given an unlabeled target bearing dataset, it is possible to make use of expert knowledge and train a data-driven fault diagnosis model for it. This is achieved by our proposed framework that combines a fault generation process and our proposed synthetic-to-real adaptation approach which is specifically designed for the imbalanced target data. 

\subsection{{Experiment Summary}}
{
One of the main findings in the experiments is that, even though the popular domain adaptation methods such as DANN and CDAN work well in the fully balanced setup, their performance can suffer significantly when class imbalance becomes larger. By providing class information to the discriminator and augment the features, we are able to alleviate this negative effect. Interestingly, we notice that the feature distribution augmentation is an important design to make the conditional alignment work well. This is likely because without the distribution augmentation, the training of the discriminator is also severely biased because of the imbalance between the classes. 
Another noteworthy finding is that even though a relatively simple model for generating the synthetic faults was chosen, it still successfully encodes the key fault information, and the proposed domain adaptation method was able to transfer the expert knowledge from the synthetic faults to a diagnosis model.}

\section{Conclusions}
We proposed a novel bearing fault diagnosis framework which can learn effective models from unlabeled real bearing data. In particular, we showed that by generating synthetic faults using expert knowledge and conducting imbalance-robust domain adaptation, a fault diagnosis model can be learned without any supervision from the real faults. We showed that a good approach to compensate the imbalance from rare target faults is key to the synthetic-to-real adaptation performance. A class conditioned adversarial adaptation method is, thus, proposed to address this issue. An additional augmentation based on the mixup approach was further proposed to deal with the limited number of fault samples and bridge the class distribution gap. The two components for domain adaptation can be easily applied in combination with other fault generation frameworks. The proposed methodology does not require any assumption on the degree of the underlying class imbalance and achieves a similar performance in the imbalanced setup as the standard DANN on the fully balanced setup, demonstrating its robustness to different imbalance levels. Experiments on the benchmark CWRU bearing dataset and a  wind turbine generator bearing dataset have validated the effectiveness of our approach also in real applications and under realistic assumptions. The framework can be implemented easily, and have the potential to be applied on other industry assets.

{
In this work, we mainly considered an unsupervised setup, where we do not have any access to the real fault labels/supervision. This is a relatively challenging task and could be unnecessarily difficult for some practical use cases. One possible direction for further improvement is to consider relaxed setups such as semi-supervised domain adaptation~\cite{yue2021prototypical}. In these setups, a few real samples for each fault type are accessible to provide additional supervision. This is often a realistic relaxation when the assets in consideration have been running for a while. We leave this for future research.}

\bibliographystyle{Bibliography/IEEEtran}
\bibliography{Bibliography/bib}

\end{document}